\documentclass[prl,twocolumn,superscriptaddress]{revtex4}

\newcommand{\ket}[1]{|{#1}\rangle}

\usepackage{amsfonts}
\usepackage{amsmath}
\usepackage{bbm}
\usepackage{graphicx}
\usepackage{bbold}
\usepackage{amssymb}
\usepackage{color}
\usepackage{subfigure}

\usepackage{mathtools}

\begin{document}
\title{On quantum time crystals and interacting gauge theories\\
in atomic Bose-Einstein condensates}

\author{Patrik \"Ohberg}
\affiliation{SUPA, Institute of Photonics and Quantum Sciences, Heriot-Watt University, Edinburgh EH14 4AS, United Kingdom}
\author{Ewan M. Wright}
\affiliation{SUPA, Institute of Photonics and Quantum Sciences, Heriot-Watt University, Edinburgh EH14 4AS, United Kingdom}
\affiliation{College of Optical Sciences, University of Arizona, Tucson, Arizona 85721, USA}

\begin{abstract}
We study the dynamics of a Bose-Einstein condensate trapped circumferentially on a ring, and which is governed by an interacting gauge theory.  We show that the associated density-dependent gauge potential and concomitant current nonlinearity permits a ground state in the form of a rotating chiral bright soliton.  This chiral soliton is constrained to move in one direction by virtue of the current nonlinearity, and represents a time crystal in the same vein as Wilczek's original proposal.
\end{abstract}

\maketitle

{\it Introduction.} A time crystal, a term first coined by Shapere and Wilczek \cite{shapere2012,wilczek2012}, is a time-periodic self organized state of a many-body system which constitutes its ground state. The original proposal for a quantum time crystal by Wilczek sparked an intense debate whether time crystals can exist or not, and arguments for and against were put forward \cite{bruno2013,nozieres2013,watanabe2015,strocchi2016}.  In particular, starting from a very general system Hamiltonian Bruno \cite{bruno2013} advanced no-go theorems that would seem to rule out Wilczek's initial proposal for a quantum time crystal.  This initial proposal was based on the mean field nonlinear Schr\"odinger equation for a system of bosons with attractive interactions trapped circumferentially on a ring and that is threaded by a flux tube.  For large enough interactions and zero flux the ground state can be well approximated by a bright soliton, so that as the flux is increased adiabatically, yielding a time-dependent flux and associated magnetic field, the bright soliton can be set in angular motion.  If the system remains in the ground state as the flux is increased the rotating bright soliton will then be a realization of a quantum time crystal.  Bruno showed that the rotating bright soliton so obtained is not the ground state but is rather a state with constant density and varying phase around the ring. In spite of the above discussion there has been a proliferation of theoretical proposals and experimental realizations of quantum time crystals that circumvent the physical assumptions underlying no-go theorems \cite{bruno2013,nozieres2013,watanabe2015}. One possibility is to use discrete and driven Floquet type systems which show dynamics that are not directly defined by the driving frequency \cite{sacha2015,else2016,khemani2016}, and this type of time crystal has been experimentally realized \cite{choi2017,zhang2017}. Another approach is to use excited states as a resource for realizing time crystals \cite{SyrZakSac17}. For a comprehensive review of time crystals we refer the reader to Ref. \cite{sacha2018}.    

In this paper we show that an atomic Bose-Einstein Condensate (BEC) which is trapped circumferentially on a ring and is governed by an interacting gauge theory can have bright soliton solutions which are chiral, i.e. the soliton can move only in one direction and is the ground state.  These chiral solitons thus represent quantum time crystals in the same vein as Wilczek's original proposal.   Whilst an atomic BEC is neutral and does not behave as a gas of charged particles in a magnetic field,  it is well established that a synthetic magnetic field can be created by coupling suitably chosen laser fields to the electronic atomic states leading the BEC to acquire an effective charge. This approach evades the radiative losses associated with using actual charged particles \cite{wilczek2012}.  The net result is an equation of motion for the BEC which includes a gauge potential determined by the applied laser parameters. Normally this gauge potential is static, meaning that it depends only on the external laser parameters, and that there is no back-action between the dynamics of the BEC and the gauge potential. For such a back-action to be present we need an interacting gauge theory.  As shown recently, this can be achieved by allowing for collisionally-induced detunings in the light-matter coupled system and leading to a density-dependent and hence nonlinear gauge potential.  This density-dependent gauge potential constitutes an interacting gauge theory and goes beyond the assumptions made in Bruno's no-go theorems, and in particular allows for chiral solitons as a realization of quantum time crystals.

{\it The interacting gauge theory.} We consider the simple setup with a BEC composed of $N$ two-level atoms with internal states  $\ket{1}$ and $\ket{2}$, which are coupled by a single laser beam. The light-matter interaction is characterized by three parameters: the Rabi frequency $\omega$ which characterizes the strength of the coupling, the phase $\phi(\mathbf{r})$ of the laser beam at the position $\mathbf{r}$ of the atom, and the detuning $\Delta$ of the laser frequency from the atomic resonance.  Then in the $\left\{\ket{1},\ket{2}\right\}$ basis the single-atom light-matter interaction is described by the operator
\begin{equation}
	U=\frac{\hbar}{2} \begin{pmatrix}
		\Delta & \omega e^{-i\phi(\mathbf{r})}\\
	 \omega e^{i\phi(\mathbf{r})} & -\Delta
	\end{pmatrix},
\label{Uaf}
\end{equation}
where $\omega$ and $\Delta$ can depend on $\bf r$. By preparing the atoms in one of the dressed states, i.e. eigenstates of the operator in Eq.\eqref{Uaf}, which we label as $\left\{\ket{+},\ket{-}\right\}$, and adiabatically eliminating the dynamics of the other component, the effective single particle Hamiltonian contains  effective vector and scalar potentials \cite{dalibard_2011}
\begin{align}
\mathbf{A}_\pm(\mathbf{r})&=-\frac{\hbar}{2}\left(1\mp\frac{\Delta}{\sqrt{\Delta^2+\omega^2}}\right)\boldsymbol{\nabla}\phi\label{A} , \\
W_\pm(\mathbf{r})&=\frac{\hbar^2}{8m}\left[\left(\boldsymbol{\nabla}\tan^{-1}\left(\frac{\omega}{\Delta}\right)\right)^2+\frac{\omega^2}{\Delta^2+\omega^2}\left(\boldsymbol{\nabla}\phi\right)^2\right] . \label{W}
\end{align}

In Ref. \cite{edmonds_2013a} it was shown how density-dependent gauge potentials arise from collision-induced detuning. At low temperatures, and in the mean field formalism, the s-wave interaction is described by the operator
\begin{equation}
	{\cal V} = \frac{\hbar}{2} \begin{pmatrix}
		\Delta_1 & 0\\
	 0 & \Delta_2
	\end{pmatrix}  ,
\label{Vaa}
\end{equation}
where $\hbar\Delta_i=g_{i1}\rho_1+g_{i2}\rho_2$ for $i=1,2$.  Here $\rho_{i}=\left|\psi_i\right|^2$ is the density of the atoms in the state $\ket{i}$, and $g_{ij}$ are the coupling constants characterizing the strength of the interaction in the different channels, with the corresponding scattering lengths $g_{ij}=4\pi\hbar^2 a_{ij}/m$. By comparing Eq. \eqref{Vaa} with \eqref{Uaf}, we see that $\Delta_1$ and $\Delta_2$ can be regarded as effective detunings induced by the collisional shift of the energy levels, and it follows that such collisions make the potentials $\mathbf{A}$ and $W$ density dependent.

We work in the weakly interacting limit, where the strength of the atom-atom interaction is much smaller than the characteristic energy of the laser-matter coupling: $\hbar\Delta_i\ll\hbar\Omega$ $(i=1,2)$ where $\Omega=\sqrt{\omega^2+\Delta^2}$ is the generalized Rabi frequency. In this limit, the interatomic interaction can be treated as a small perturbation to the light-atom interaction. Also, we work with the simplest set-up, in which the laser field is perfectly resonant with the atomic transition, so that $\Delta=0$, and the Rabi frequency is homogeneous in space. Then keeping terms up to first-order the gauge potential becomes
\begin{equation}
	\mathbf{A}_\pm =\mathbf{A}^{(0)}\pm \mathbf{a}_1 |\Psi_\pm|^2, \label{Vec_Pot}
\end{equation}
with $\mathbf{A}^{(0)}=-\frac{\hbar}{2}\nabla\phi$ the single particle component of the vector potential,  and $W =|\mathbf{A}^{(0)}|^2/{2m}$  the scalar potential, while the vector field $\mathbf{a}_1=\boldsymbol{\nabla}\phi\,(g_{11}-g_{22})/8\Omega$ controls the strength of the first-order density-dependent contribution to the vector potential. 

By minimizing the Dirac-Frenkel action \cite{butera2015quantized}
\begin{equation}
	S_\pm=\int{dtd\mathbf{r}~\Psi_\pm^*\left(i\hbar{\partial \over \partial t}-H_\pm\right)\Psi_\pm }  
\label{Action}
\end{equation}
with respect to $\Psi_\pm^*$, where
\begin{equation}
	H_\pm=\frac{\left(\mathbf{p}-\mathbf{A}_\pm\right)^2}{2m}+W+\frac{g}{2} |\Psi_\pm|^2 
\label{Hamil_pm}
\end{equation}
is the effective mean field Hamiltonian with $g=(g_{11}+g_{22}-2g_{12})/4$, we obtain the Gross-Pitaevskii equation (GPE) for the order parameter
\begin{equation}
	i\hbar\frac{\partial\Psi_\pm}{\partial t}=\left[\frac{\left(\mathbf{p}-\mathbf{A}_\pm\right)^2}{2m}\mp \mathbf{a}_1\cdot \mathbf{j}_\pm +W+g|\Psi_\pm|^2 \right]\Psi_\pm  ,
\label{GP}
\end{equation}
with the current nonlinearity defined as
\begin{equation}
	\mathbf{j}_\pm=\frac{\hbar}{2mi} \Psi_\pm^*\left(\boldsymbol{\nabla}-\frac{i}{\hbar} \mathbf{A}_\pm\right)\Psi_\pm + c.c . \label{Current}
\end{equation}
In the following we consider the $(+)$ component of the condensate, and hereafter drop the subscript in the quantities defined above for brevity in notation.
%
%

To proceed we specialize to the one-dimensional case and recast Eq.~\eqref{GP} as
\begin{equation}\label{EOM}
i\hbar\frac{\partial\psi}{\partial t}=\bigg[-\frac{\hbar^2}{2m}{\partial^2\over\partial x^2}+W-2a_1Nj(x)+gN\vert\psi\vert^2\bigg]\psi,
\end{equation}
with spatial coordinate $x$, and current nonlinearity
\begin{equation}\label{jprime}
j(x)=\frac{\hbar}{2mi}\left(\psi^*{\partial\psi\over\partial x}-\psi{\partial\psi^*\over\partial x} \right),
\end{equation}
which is arrived at by using the nonlinear transformation 
\begin{equation}
\Psi\left(x,t\right)=\sqrt{N}\psi\left(x,t\right)e^{-i\frac{\phi}{2}+\frac{ia_1}{\hbar}\int^x_{\infty}dx^\prime\;\vert\psi\left(x^\prime,t\right)\vert^2},
\end{equation}
where $N$ is the number of atoms in the condensate.  In the literature, Eq.~\eqref{EOM} is often referred to as a chiral nonlinear Schr\"{o}dinger equation, which was originally studied in the context of one-dimensional anyons \cite{aglietti_1996}. 

In anticipation of looking for solitons that are moving we introduce the Galilean transformation
\begin{equation}\label{trans_Galil}
\psi(x,t)=\Phi(x^\prime,t^\prime)e^{i(mux^{\prime}+mu^{2}t^{\prime
}/2)/\hbar },
\end{equation}
where the stationary coordinates $(x,t)$ and moving coordinates $(x^\prime,t^\prime)$ are related by the translations, $x^\prime\rightarrow (x-ut)$ and $t^\prime\rightarrow t$, with frame velocity $u$. The dynamics of the condensate in the moving frame is then described by the equation of motion
\begin{equation}\label{EOM_mf}
i\hbar\frac{\partial\Phi}{\partial t^\prime}=\bigg[-\frac{\hbar^2}{2m}{\partial^2\over\partial x^2} +W-2a_1Nj(x^\prime)+\left(g -2a_1u\right)N\vert\Phi\vert^2\bigg]\Phi.
\end{equation}
For simplicity in notation we hereafter explicitly drop the prime notation and work in the moving frame unless otherwise stated.  The chiral bright soliton solutions of Eq.~\eqref{EOM_mf} on the infinite line, with zero boundary conditions on the wave function and its derivative at $x=\pm\infty$, then admit the standard form $\Phi\left(x,t\right)=\chi(x)e^{-i\mu t/\hbar}$ for $\tilde{g}=(g-2a_1u)<0$, with spatial profile
\begin{equation}\label{brightsoliton}
\chi\left(x\right)={\frac{1}{\sqrt{2b}}} {1\over\cosh \left ( x/b \right )} ,
\end{equation}
the width $b=-2\hbar^2/m\tilde{g}N$, and chemical potential $\mu=-m\tilde{g}^2N^2/8\hbar^2$.  In addition, both the width and chemical potential of the soliton depend on the direction of motion dictated by the sign of the velocity $u$, this being a result of the breakdown of Galilean invariance of Eq.~\eqref{EOM}.

\begin{figure}
\includegraphics[width=8.5cm]{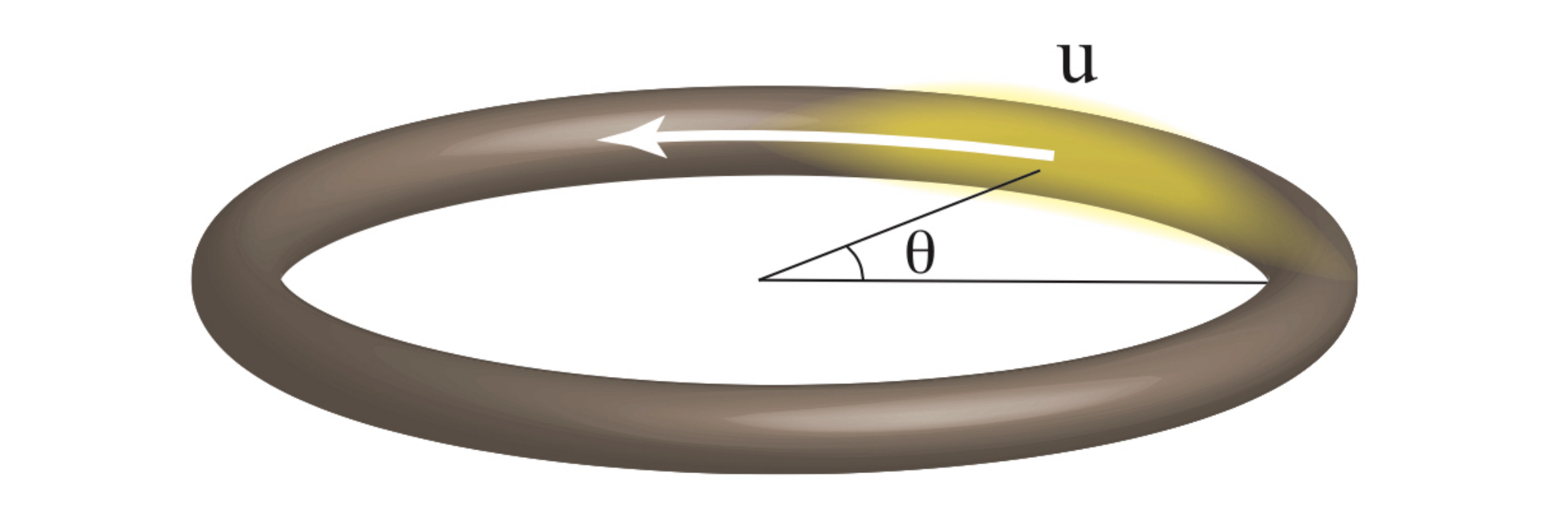}
\caption{The tightly confined ring with radius $R$ circumferentially traps the Bose-Einstein condensate such that the dynamics is effectively one-dimensional along the azimuthal or $\theta$-direction. The chiral soliton of velocity $u$, indicated by the yellow region, is only allowed to travel in one specific direction given by the parameters of the current-nonlinearity.}
\label{figexp}
\end{figure}

{\it Rotating chiral solitons on a ring.} Equation \eqref{EOM_mf} represents an interacting gauge theory for the BEC on the infinite line with density-dependent gauge potentials and associated current nonlinearity. Next we consider a ring trap with radius $R$ in which the atoms are tightly confined in the radial direction and the dynamics is effectively one-dimensional along the azimuthal direction $\theta$, see Fig. \ref{figexp} for an illustration of this geometry.  We may apply Eq. \eqref{EOM} to this situation if we identify the coordinate $x$ as parametrizing the arc length around the ring $x=R\theta$, with the caveat that periodic boundary conditions must be applied to the wave function $\psi(x,t)=\psi(x+2\pi R,t)$, or $\psi(\theta,t)=\psi(\theta+2\pi,t)$: This is in stark contrast to the boundary conditions at $x=\pm\infty$ applied to the bright soliton on the infinite line given in Eq. (\ref{brightsoliton}).  Our goal is to explore whether chiral solitons obeying these periodic boundary conditions can attain their ground state whilst being in a state of rotation. If such chiral soliton solutions exist they are candidates for realizing quantum time crystals.

For the gauge potential in Eq. (\ref{Vec_Pot}) we choose an incident Laguerre-Gaussian (LG) laser beam with orbital angular momentum $\hbar q$ per photon where $q=0,\pm 1,\pm 2,\ldots$. The gauge potential is then in the azimuthal $\theta$-direction where the zeroth-order contribution is given by ${\bf A}_0=\hbar q/(2R)$, and $a_1=q(g_{11}-g_{22})/8R\Omega$ controls the strength of the current nonlinearity, both of these depending on the winding number $q$ of the LG laser beam.

To highlight the general nature of our results we transfer to dimensionless form by expressing all length scales in units of the ring radius $R$, $\theta=x/R$, time $\tau$ in units of $2mR^2/\hbar$, energy in units of $\hbar^2/2mR^2$, and replace $\Phi(x,t)\rightarrow \varphi(\theta,\tau)$.  Then Eq.~\eqref{EOM_mf} for the dimensionless order parameter, expressed in the frame rotating at (dimensionless) velocity $u$, becomes
\begin{equation}
	i{\partial\varphi(\theta,\tau)\over\partial \tau} =\left [-{\partial^2\over\partial\theta^2} -2a j(\varphi)+\tilde{\textsl{g}}|\varphi |^2 \right ]\varphi(\theta,\tau) ,
\label{GPS}
\end{equation}
with norm ${\cal N}=\int_0^{2\pi} d\theta |\varphi(\theta,\tau)|^2=1$.  Here the scaled current nonlinearity is given by $j(\varphi)={\rm Im}\left [\varphi^* {\partial\varphi\over\partial\theta} \right ]$,
where the dimensionless constant $a=a_1 N/\hbar$ characterizes the strength of the current nonlinearity, and $\tilde{\textsl{g}}=(\textsl{g}-2au)$ is the dimensionless form of $N\tilde g$.  For the first-order approximation to the vector potential in Eq. (\ref{Vec_Pot}) to be valid then requires that $|{a\over q}|=|{(g_{11}-g_{22})N/R\over 4\hbar \Omega}|\ll1$.

We are interested in stationary solutions of Eq.~\eqref{GPS} of the form $\varphi(\theta,\tau)=e^{-i\beta\tau}\chi(\theta)$ where 
\begin{equation}
	\beta\chi=\left [-{d^2\over d\theta^2} -2a j(\chi)+\tilde{\textsl{g}}|\chi |^2 \right ]\chi  ,
\label{GPS_TI}
\end{equation}
$\beta$ being the scaled chemical potential.  Analytic solutions of Eq.~\eqref{GPS_TI} with periodic boundary conditions and $j=0$ are given in terms of Jacobi elliptic functions in Refs. \cite{Carr2000} and \cite{Kanamoto2003}.  We have verified that for $\tilde{\textsl{g}} < 0$ these Jacobi elliptic function solutions coincide with the ground states found by numerically propagating Eq.~\eqref{GPS} in imaginary time starting from an initial narrow Gaussian wavepacket that can have $j\ne 0$. For large enough times the imaginary time propagation method relaxes towards the ground state which is stable by virtue of having the lowest chemical potential \cite{LehToiElo2007}, and we found that these have $j=0$.  The Jacobi elliptic function solutions therefore represent rotating chiral solitons: In conjunction with the fact that these have $j=0$, the periodic boundary conditions for our ring geometry lead to quantized scaled velocities and associated angular momenta.  This can be deduced by examining the spatially varying term from the Galilean transformation in Eq. \eqref{trans_Galil}, such that $e^{imux/\hbar} \rightarrow e^{iu\theta/2}$, where scaled variables appear on the right-hand-side.  Based on this we see that in order for the associated wave functions to be single-valued we require that $u=0,\pm1,\pm2,\ldots$ be an even integer.

\begin{figure}
\includegraphics[width=1.1\linewidth,clip]{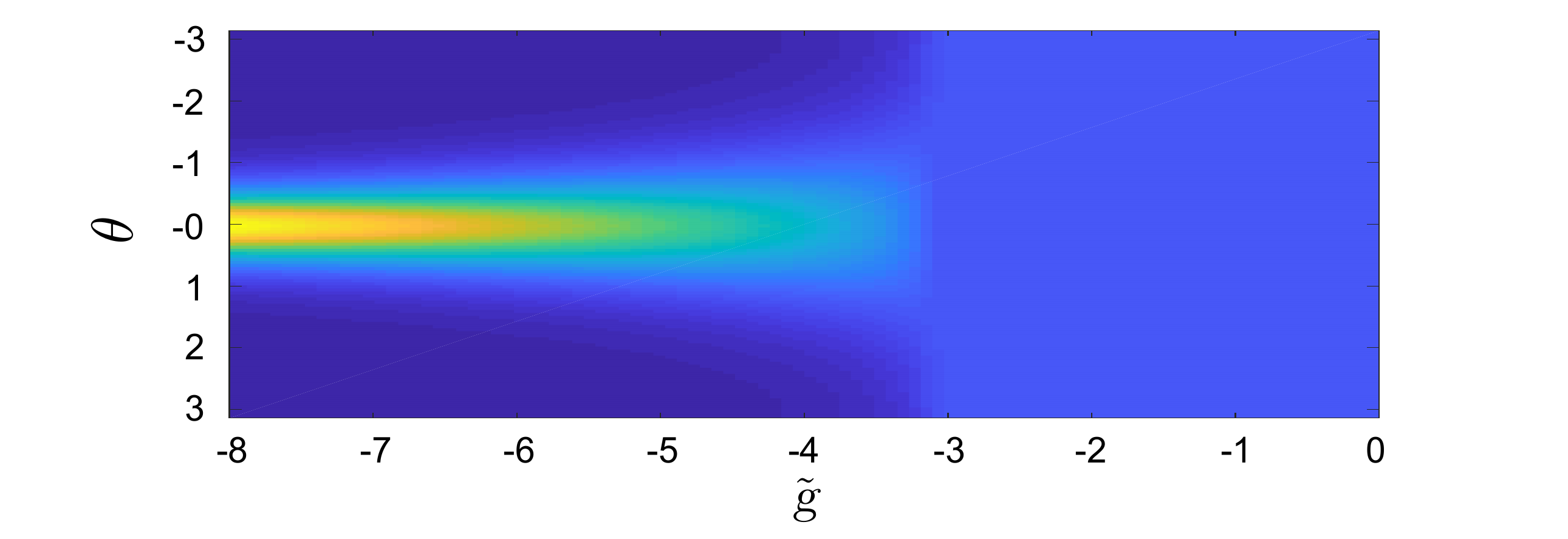}
\caption{Color coded plot of the chiral soliton density profile $|\chi(\theta)|^2$ obtained from the numerical solutions versus the angular variable $\theta$ and $\tilde{\textsl{g}}$.} \label{Fig2}
\end{figure}

Figure \ref{Fig2} shows a color coded plot of the chiral soliton density profile $|\chi(\theta)|^2$ obtained from the numerical solutions versus the angular variable $\theta$ and for a range of values of $\tilde{\textsl{g}}<0$.  We note that the chiral soliton density profile is homogeneous for $\tilde{\textsl{g}}> -\pi$ and becomes progressively more inhomogeneous for $\tilde{\textsl{g}} < -\pi$.  This is in accordance with the results of Kanamoto et al. \cite{Kanamoto2003} who showed that a quantum phase transition from a homogeneous state towards a localized soliton state occurs at $\tilde{\textsl{g}}=-\pi$ as adapted to our notation.  With reference to Fig. \ref{Fig2} we note that as $\tilde{\textsl{g}}<-\pi$ becomes more negative the density profile becomes narrower than the $2\pi$ angular extent of the ring.   In this limit the density profile approaches that of the bright soliton in Eq.~\eqref{brightsoliton} which in our scaled units takes the form
\begin{equation}\label{CSapp}
\chi(\theta)={\frac{1}{\sqrt{2b}}} {1\over\cosh \left ( \theta/b \right )} , \quad b={4\over |\tilde{\textsl{g}}|} .
\end{equation}
Transforming back to the laboratory frame the stationary solutions yield the scaled current $j(\theta,\tau)={u\over 2}|\chi(\theta-u\tau)|^2$, a localized and rotating current that could be observed.

{\it The quantum time crystal.} So far we have established that rotating chiral solitons can arise from the interacting gauge theory, but to claim that these can be used to realize a quantum time crystal requires that such solitons can also be ground states of lowest energy under suitable conditions.  For this purpose we need to identify a suitable energy functional: In the rotating frame with $j=0$ this functional is
\begin{equation}
E' = \int_0^{2\pi} d\theta \left ( \left |{d\chi\over d\theta} \right |^2 +{\tilde{\textsl{g}}\over 2} |\chi |^4 \right )  .
\end{equation}
With this energy functional the time-independent GPE~\eqref{GPS_TI} follows from extremizing the functional $(E'-\beta {\cal N})$ with respect to variation in $\chi^*$.  However, we need the energy in the non-rotating lab frame, and for this purpose we must account for the phase factor $e^{iu\theta/2}$ that arises from the Galilean transformation.  The energy in the lab frame then becomes
\begin{equation}
E = {u^2\over 4} + \int_0^{2\pi} d\theta \left ( \left |{d\chi\over d\theta}\right |^2 +{\tilde{\textsl{g}}\over 2} |\chi |^4 \right )  .
\end{equation}
Using the approximate chiral soliton solution in Eq. (\ref{CSapp}) for $\tilde{\textsl{g}}<-\pi$, and integrating the energy over the infinite line, and using the homogeneous solution for $\tilde{\textsl{g}}>-\pi$, yields the approximation to the energy
\begin{equation}\label{Eapp}
E = \left \{ 
\begin{array}{c}
{u^2\over 4} - {\tilde{\textsl{g}}^2\over 48} , \quad \tilde{\textsl{g}}<-\pi \\
{u^2\over 4} + {\tilde{\textsl{g}}\over 4\pi } ,\quad \tilde{\textsl{g}}> -\pi  .
\end{array}
\right .
\end{equation}
For $\tilde{\textsl{g}}<-\pi$, so that a bright chiral soliton is formed, a consequence of this form of the energy is that in order to obtain an energy minimum with respect to the scaled velocity $u$, with ${\partial^2 E\over\partial u^2}> 0$, requires $0<|a|<\sqrt{3}=1.732$.  The energy minimum occurs then for $u_{min}=-\left ({1\over 2} \right ){a\textsl{g}\over 3-a^2}$.

\begin{figure}
\includegraphics[width=1.1\linewidth,clip]{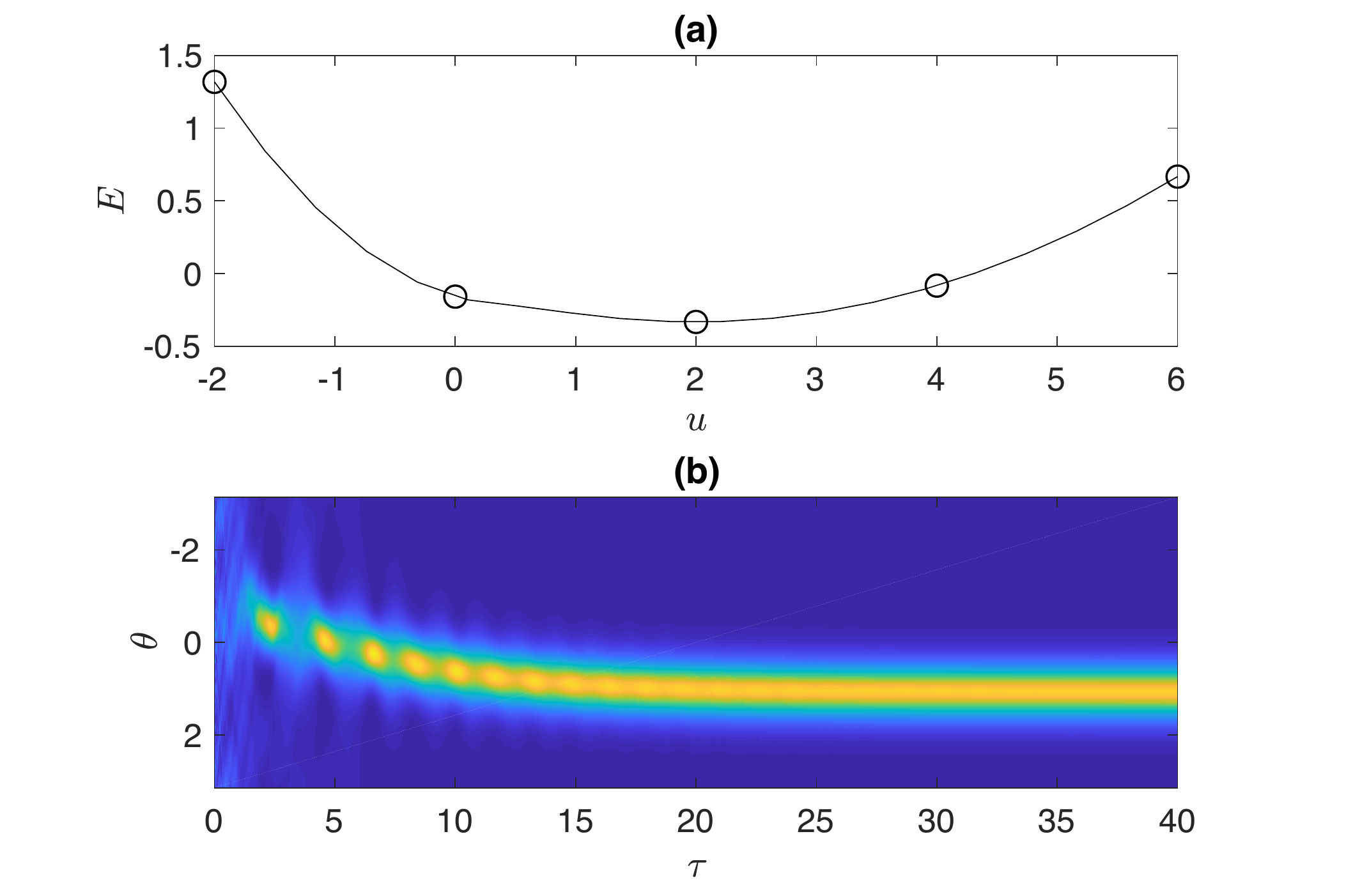}
\caption{(a) Numerically calculated scaled energy $E$ versus the scaled velocity $u$ for $a=1.5$ and $\textsl{g}=-2$,  showing an energy minimum for $u=2$.  The circles provide the approximate energy in Eq. (\ref{Eapp}) for the allowed quantized values of the velocity $u=0,\pm 2,\ldots$. (b) Color coded plot of the time evolution of the probability density $|\varphi(\theta,\tau)|^2$ for $\Lambda=-0.05$, and starting from a random initial condition.}  
\label{Fig3}
\end{figure}

The key to realizing a quantum time crystal is to include $\textsl{g}\ne 0$ so that $u_{min}$ above can be non-zero.  An example is shown in Fig. \ref{Fig3}(a) which shows the scaled energy $E$ versus scaled velocity $u$ for $a=1.5$ and $\textsl{g}=-2$ which yields $u_{min}=2$, one of the allowed quantized velocities (the circles show the energy according to the approximate expression in Eq. (\ref{Eapp})).  Furthermore, for this example $\tilde{\textsl{g}}=-8$, so that the corresponding chiral soliton is spatially localized, see Fig. (\ref{Fig2}).  This provides an explicit demonstration that our proposed system can be used to produce a ground state that is a rotating and bright chiral soliton, that is, a quantum time crystal.  More generally, for a given value of the parameter $'a'$ and choice of $u$ equal to one of the allowed quantized values, the expression for $u_{min}$ can be used to calculate the required value of $\textsl{g}$.  If this also leads to $\tilde{\textsl{g}}<-\pi$ then a quantum time crystal will be realized, as long as the parameters are within the limits of validity of the theory.

To illustrate the robustness of the rotating chiral soliton above we used the phenomenological model for damping in an atomic BEC given in Ref. \cite{ChoMorBur98} to simulate the formation of the quantum time crystal.  In the rotating frame the damped GPE generalizing Eq. (\ref{GPS}) becomes
\begin{equation}
	i{\partial\varphi\over\partial \tau} =(1+i\Lambda)\left [-{\partial^2\over\partial\theta^2} -2a j(\varphi)+\tilde{\textsl{g}}|\varphi |^2 \right ]\varphi ,
\end{equation}
where $\Lambda<0$ is inversely proportional to the (scaled) relaxation time: This relaxation can arise, for example, from the interaction between condensate and thermally excited atoms.  Figure \ref{Fig3}(b) shows a color coded plot of the time $(\tau)$ evolution of the probability density $|\varphi(\theta,\tau)|^2$ for $\Lambda=-0.05$, and starting from a random initial condition.  Here we see that even for such an initial condition that is very far from the ground state, the solution evolves in time $\tau$ towards the rotating chiral soliton (that is stationary in the frame rotating at scaled velocity $u$).  Although not all-encompassing this simulation indicates that precise initial conditions are not required to realize the quantum time crystal.
 
{\it Discussion.} The results presented in this section clearly show that a quantum time crystal can be realized for our example of an atomic Bose-Einstein condensate trapped circumferentially on a ring.  This conclusion apparently runs counter to previous no-go theorems, due to Bruno \cite{bruno2013}, forbidding rotating and localized ground states in a general setting.  It therefore behooves us to elucidate why the no-go theorems do not apply to our system, and there are two ways to phrase this:  First, Bruno considered only the case of static gauge potentials, whereas here we have an interacting gauge theory with density-dependent potentials.  Alternatively, Bruno accounted for cubic nonlinearities as in the standard nonlinear Schr\"odinger equation \cite{BrunoComment}, but the interacting gauge theory is naturally described by a derivative nonlinear Schr\"odinger equation with current coupling \cite{edmonds_2013a}.  For these reasons the no-go theorems of Bruno do not apply to our model and this allow for a quantum time crystal close in spirit to Wilczek's original proposal.

{\it Conclusions.} In summary, we have shown that a quantum time crystals akin to those envisioned by Wilczek can be realized using an interacting gauge theory for a Bose-Einstein condensate of atoms on a ring.  The key new ingredient that has made this possible is the appearance of density-dependent gauge potentials that were absent from previous considerations. This can also be seen as a variant of the derivative nonlinear Schr\"odinger equation due to the current-nonlinearity. The corresponding quantum many-body Hamiltonian also allows for bound states, which provides an intriguing link between the mean-field and the fully quantum many-body situation \cite{PhysRevA.40.844,PhysRevA.50.3453,sen1996}. Furthermore, motivated by Refs. \cite{Kanamoto} and \cite{Javan} we anticipate that mean-field rotating chiral solitons can emerge via the measurement process applied to the many-body system even for finite atom number N, without reliance on the thermodynamic limit that may not apply to small atomic BECs. The experimental realization of this proposal would constitute a new type of quantum time crystal that does not rely on a discrete system or use of an exited state to attain periodic time variation. 

\begin{acknowledgements} 
We thank Brian Anderson, Yvan Buggy, Kieran Fraser, Rina Kanamoto and Manuel Valiente for useful discussions. We also thank an anonymous reviewer for very constructive input. P.\"O acknowledges support from EPSRC EP/M024636/1.  
\end{acknowledgements}


\begin{thebibliography}{16}
\expandafter\ifx\csname natexlab\endcsname\relax\def\natexlab#1{#1}\fi
\expandafter\ifx\csname bibnamefont\endcsname\relax
  \def\bibnamefont#1{#1}\fi
\expandafter\ifx\csname bibfnamefont\endcsname\relax
  \def\bibfnamefont#1{#1}\fi
\expandafter\ifx\csname citenamefont\endcsname\relax
  \def\citenamefont#1{#1}\fi
\expandafter\ifx\csname url\endcsname\relax
  \def\url#1{\texttt{#1}}\fi
\expandafter\ifx\csname urlprefix\endcsname\relax\def\urlprefix{URL }\fi
\providecommand{\bibinfo}[2]{#2}
\providecommand{\eprint}[2][]{\url{#2}}

\bibitem[{\citenamefont{Shapere and Wilczek}(2012)}]{shapere2012}
\bibinfo{author}{\bibfnamefont{A.}~\bibnamefont{Shapere}} \bibnamefont{and}
  \bibinfo{author}{\bibfnamefont{F.}~\bibnamefont{Wilczek}},
  \bibinfo{journal}{Phys. Rev. Lett.} \textbf{\bibinfo{volume}{109}},
  \bibinfo{pages}{160402} (\bibinfo{year}{2012}).

\bibitem[{\citenamefont{Wilczek}(2012)}]{wilczek2012}
\bibinfo{author}{\bibfnamefont{F.}~\bibnamefont{Wilczek}},
  \bibinfo{journal}{Phys. Rev. Lett.} \textbf{\bibinfo{volume}{109}},
  \bibinfo{pages}{160401} (\bibinfo{year}{2012}).

\bibitem[{\citenamefont{Bruno}(2013)}]{bruno2013}
\bibinfo{author}{\bibfnamefont{P.}~\bibnamefont{Bruno}},
  \bibinfo{journal}{Phys. Rev. Lett.} \textbf{\bibinfo{volume}{111}},
  \bibinfo{pages}{070402} (\bibinfo{year}{2013}).

\bibitem[{\citenamefont{Nozières}(2013)}]{nozieres2013}
\bibinfo{author}{\bibfnamefont{P.}~\bibnamefont{Nozières}},
  \bibinfo{journal}{EPL (Europhysics Letters)} \textbf{\bibinfo{volume}{103}},
  \bibinfo{pages}{57008} (\bibinfo{year}{2013}).

\bibitem[{\citenamefont{Watanabe and Oshikawa}(2015)}]{watanabe2015}
\bibinfo{author}{\bibfnamefont{H.}~\bibnamefont{Watanabe}} \bibnamefont{and}
  \bibinfo{author}{\bibfnamefont{M.}~\bibnamefont{Oshikawa}},
  \bibinfo{journal}{Phys. Rev. Lett.} \textbf{\bibinfo{volume}{114}},
  \bibinfo{pages}{251603} (\bibinfo{year}{2015}).

\bibitem[{\citenamefont{Strocchi}(2016)}]{strocchi2016}
\bibinfo{author}{\bibfnamefont{F.}~\bibnamefont{Strocchi}} \bibnamefont{and}
  \bibinfo{author}{\bibfnamefont{C.}~\bibnamefont{Heissenberg}},
  \bibinfo{journal}{arXiv preprint arXiv:1605.04188}  (\bibinfo{year}{2016}).

\bibitem[{\citenamefont{Sacha}(2015)}]{sacha2015}
\bibinfo{author}{\bibfnamefont{K.}~\bibnamefont{Sacha}},
  \bibinfo{journal}{Phys. Rev. A} \textbf{\bibinfo{volume}{91}},
  \bibinfo{pages}{033617} (\bibinfo{year}{2015}).

\bibitem[{\citenamefont{Else et~al.}(2016)\citenamefont{Else, Bauer, and
  Nayak}}]{else2016}
\bibinfo{author}{\bibfnamefont{D.~V.} \bibnamefont{Else}},
  \bibinfo{author}{\bibfnamefont{B.}~\bibnamefont{Bauer}}, \bibnamefont{and}
  \bibinfo{author}{\bibfnamefont{C.}~\bibnamefont{Nayak}},
  \bibinfo{journal}{Phys. Rev. Lett.} \textbf{\bibinfo{volume}{117}},
  \bibinfo{pages}{090402} (\bibinfo{year}{2016}).

\bibitem[{\citenamefont{Khemani et~al.}(2016)\citenamefont{Khemani, Lazarides,
  Moessner, and Sondhi}}]{khemani2016}
\bibinfo{author}{\bibfnamefont{V.}~\bibnamefont{Khemani}},
  \bibinfo{author}{\bibfnamefont{A.}~\bibnamefont{Lazarides}},
  \bibinfo{author}{\bibfnamefont{R.}~\bibnamefont{Moessner}}, \bibnamefont{and}
  \bibinfo{author}{\bibfnamefont{S.~L.} \bibnamefont{Sondhi}},
  \bibinfo{journal}{Phys. Rev. Lett.} \textbf{\bibinfo{volume}{116}},
  \bibinfo{pages}{250401} (\bibinfo{year}{2016}).

\bibitem[{\citenamefont{Choi et~al.}(2017)\citenamefont{Choi, Choi, Landig,
  Kucsko, Zhou, Isoya, Jelezko, Onoda, Sumiya, Khemani et~al.}}]{choi2017}
\bibinfo{author}{\bibfnamefont{S.}~\bibnamefont{Choi}},
  \bibinfo{author}{\bibfnamefont{J.}~\bibnamefont{Choi}},
  \bibinfo{author}{\bibfnamefont{R.}~\bibnamefont{Landig}},
  \bibinfo{author}{\bibfnamefont{G.}~\bibnamefont{Kucsko}},
  \bibinfo{author}{\bibfnamefont{H.}~\bibnamefont{Zhou}},
  \bibinfo{author}{\bibfnamefont{J.}~\bibnamefont{Isoya}},
  \bibinfo{author}{\bibfnamefont{F.}~\bibnamefont{Jelezko}},
  \bibinfo{author}{\bibfnamefont{S.}~\bibnamefont{Onoda}},
  \bibinfo{author}{\bibfnamefont{H.}~\bibnamefont{Sumiya}},
  \bibinfo{author}{\bibfnamefont{V.}~\bibnamefont{Khemani}},
  \bibnamefont{et~al.}, \bibinfo{journal}{Nature}
  \textbf{\bibinfo{volume}{543}}, \bibinfo{pages}{221} (\bibinfo{year}{2017}).

\bibitem[{\citenamefont{Zhang et~al.}(2017)\citenamefont{Zhang, Hess,
  Kyprianidis, Becker, Lee, Smith, Pagano, Potirniche, Potter, Vishwanath
  et~al.}}]{zhang2017}
\bibinfo{author}{\bibfnamefont{J.}~\bibnamefont{Zhang}},
  \bibinfo{author}{\bibfnamefont{P.~W.} \bibnamefont{Hess}},
  \bibinfo{author}{\bibfnamefont{A.}~\bibnamefont{Kyprianidis}},
  \bibinfo{author}{\bibfnamefont{P.}~\bibnamefont{Becker}},
  \bibinfo{author}{\bibfnamefont{A.}~\bibnamefont{Lee}},
  \bibinfo{author}{\bibfnamefont{J.}~\bibnamefont{Smith}},
  \bibinfo{author}{\bibfnamefont{G.}~\bibnamefont{Pagano}},
  \bibinfo{author}{\bibfnamefont{I.-D.} \bibnamefont{Potirniche}},
  \bibinfo{author}{\bibfnamefont{A.~C.} \bibnamefont{Potter}},
  \bibinfo{author}{\bibfnamefont{A.}~\bibnamefont{Vishwanath}},
  \bibnamefont{et~al.}, \bibinfo{journal}{Nature}
  \textbf{\bibinfo{volume}{543}}, \bibinfo{pages}{217} (\bibinfo{year}{2017}).
  
  \bibitem{SyrZakSac17}
  A. Syrwid, J. Zakrzewski, and K. Sacha, {\sl Phys. Rev. Lett.} {\bf 119}, 250602 (2017).

\bibitem[{\citenamefont{Sacha and Zakrzewski}(2018)}]{sacha2018}
\bibinfo{author}{\bibfnamefont{K.}~\bibnamefont{Sacha}} \bibnamefont{and}
  \bibinfo{author}{\bibfnamefont{J.}~\bibnamefont{Zakrzewski}},
  \bibinfo{journal}{Reports on Progress in Physics}
  \textbf{\bibinfo{volume}{81}}, \bibinfo{pages}{016401}
  (\bibinfo{year}{2018}).

\bibitem[{\citenamefont{Dalibard et~al.}(2011)\citenamefont{Dalibard, Gerbier,
  Juzeli\ifmmode~\bar{u}\else \={u}\fi{}nas, and \"Ohberg}}]{dalibard_2011}
\bibinfo{author}{\bibfnamefont{J.}~\bibnamefont{Dalibard}},
  \bibinfo{author}{\bibfnamefont{F.}~\bibnamefont{Gerbier}},
  \bibinfo{author}{\bibfnamefont{G.}~\bibnamefont{Juzeli\ifmmode~\bar{u}\else
  \={u}\fi{}nas}}, \bibnamefont{and}
  \bibinfo{author}{\bibfnamefont{P.}~\bibnamefont{\"Ohberg}},
  \bibinfo{journal}{Rev. Mod. Phys.} \textbf{\bibinfo{volume}{83}},
  \bibinfo{pages}{1523} (\bibinfo{year}{2011}).

\bibitem[{\citenamefont{Edmonds et~al.}(2013)\citenamefont{Edmonds, Valiente,
  Juzeli\=unas, Santos, and \"Ohberg}}]{edmonds_2013a}
\bibinfo{author}{\bibfnamefont{M.~J.} \bibnamefont{Edmonds}},
  \bibinfo{author}{\bibfnamefont{M.}~\bibnamefont{Valiente}},
  \bibinfo{author}{\bibfnamefont{G.}~\bibnamefont{Juzeli\=unas}},
  \bibinfo{author}{\bibfnamefont{L.}~\bibnamefont{Santos}}, \bibnamefont{and}
  \bibinfo{author}{\bibfnamefont{P.}~\bibnamefont{\"Ohberg}},
  \bibinfo{journal}{Phys. Rev. Lett.} \textbf{\bibinfo{volume}{110}},
  \bibinfo{pages}{085301} (\bibinfo{year}{2013}).

\bibitem[{\citenamefont{Butera et~al.}(2015)\citenamefont{Butera, Valiente, and
  {\"O}hberg}}]{butera2015quantized}
\bibinfo{author}{\bibfnamefont{S.}~\bibnamefont{Butera}},
  \bibinfo{author}{\bibfnamefont{M.}~\bibnamefont{Valiente}}, \bibnamefont{and}
  \bibinfo{author}{\bibfnamefont{P.}~\bibnamefont{{\"O}hberg}},
  \bibinfo{journal}{Journal of Physics B: Atomic, Molecular and Optical
  Physics} \textbf{\bibinfo{volume}{49}}, \bibinfo{pages}{015304}
  (\bibinfo{year}{2015}).

\bibitem[{\citenamefont{Aglietti et~al.}(1996)\citenamefont{Aglietti, Griguolo,
  Jackiw, Pi, and Seminara}}]{aglietti_1996}
\bibinfo{author}{\bibfnamefont{U.}~\bibnamefont{Aglietti}},
  \bibinfo{author}{\bibfnamefont{L.}~\bibnamefont{Griguolo}},
  \bibinfo{author}{\bibfnamefont{R.}~\bibnamefont{Jackiw}},
  \bibinfo{author}{\bibfnamefont{S.~Y.} \bibnamefont{Pi}}, \bibnamefont{and}
  \bibinfo{author}{\bibfnamefont{D.}~\bibnamefont{Seminara}},
  \bibinfo{journal}{Phys. Rev. Lett.} \textbf{\bibinfo{volume}{77}},
  \bibinfo{pages}{\href{http://journals.aps.org/prl/abstract/10.1103/PhysRevLett.77.4406}{4406}}
  (\bibinfo{year}{1996}).
  
  \bibitem{Carr2000}
  L. D. Carr, C. W. Clark, and W. P. Reinhardt, {\sl Phys. Rev. A} {\bf 62}, 063611 (2000).
  
  \bibitem{Kanamoto2003}
  R. Kanamoto, H. Saito, and M. Ueda, {\sl Phys. Rev. A} {\bf 67}, 013608 (2003).
  
  \bibitem{LehToiElo2007}
  See, for example, L. Lehtovaara, J. Toivanen, and J. Eloranta, {\sl J. of Comp. Phys.} {\bf 221}, 148 (2007).
  
  \bibitem{ChoMorBur98}
  S. Choi, S. A. Morgan, and K. Burnett, {\sl Phys. Rev. A} {\bf 57}, 4057 (1998).
  
  \bibitem{BrunoComment}
  P. Bruno, {\sl Phys. Rev. Lett.} {\bf 110}, 118901 (2013).

\bibitem{PhysRevA.40.844}
Y. Lai and H. A. Haus, {\sl Phys. Rev. A} {\bf 40}, 844 (1989).


\bibitem{PhysRevA.50.3453}
  A. G. Shnirman, B. A. Malomed, and B.-J Eshel, {\sl Phys. Rev. A} {\bf 40}, 3453 (1994).

\bibitem{sen1996}
D. Sen, {\sl arXiv:cond-mat/9612077}.

\bibitem{Kanamoto}
R. Kanamoto, H. Saito, and M. Ueda, {\sl Phys. Rev. A} {\bf 73}, 033611 (2006).

\bibitem{Javan}
J. Javanainen and U. Shrestha, {\sl Phys. Rev. Lett.} {\bf 101}, 170405 (2008).

\end{thebibliography}

\end{document}